\documentclass{jjap3}

\title{Stacking sequence dependence of electronic properties in
double-layer graphene heterostructures }

\author{Mei-Yan Ni$^{1,2}$\thanks{E-mail address: nimy@hfut.edu.cn} and Katsunori Wakabayashi$^{2}$
\thanks{E-Mail address: WAKABAYASHI.Katsunori@nims.go.jp}}

\inst{$^{1}$School of Electronic Science and Applied Physics, Hefei
University of Technology, Hefei 230009, P.R. China \\
$^{2}$International Center for Materials Nanoarchitectonics (WPI-MANA),
National Institute for Materials Science (NIMS), Namiki 1-1, Tsukuba 305-0044, Japan}

\abst{
First-principles calculation has been performed to investigate
the stability and electronic properties of double-layer graphene
heterostructure (DLGH). In this system, two graphene layers are separated by
hexagonal boron-nitride (h-BN) layers which work as a insulating barrier.
Our results show that the stability of the system is determined by the
atomistic configurations between graphene and its adjacent h-BN
layer. Among different stacking sequences, Ab-stacking is most stable.
Since the inserted h-BN layers modulate the on-site
energies for carbon atoms of graphene layers,
the electronic states of the system can be classified into metallic or
semiconducting by the stacking sequence.
And the stacking sequence dependence of the energy band structures of DLGHs
are well described by the orbital interaction model.
}

\begin{document}
\maketitle

\section{Introduction}
Monolayer graphene invokes great interest in electronic application
owing to the unique properties such as ultra-high electronic mobility,
anomalous quantum Hall effects and so on.~\cite {KSNovoselov1,YBZhang1,
AKGeim1,KIBolotin,AABalandin,TAndo,FSchweirz}
These unconventional electronic properties are originated from the unusual
band structure near the Fermi level of graphene, where the electronic
states are well described as massless Dirac electrons, i.e.
a zero energy gap with linear $\pi$ bands crossing at the $K$ point.
In few-layer graphene system, however, the weak interaction between graphene layers
changes the electronic properties near the Fermi level.~\cite{MAoki,MKlintenberg, JYan, FZhang, DHXu}
For bilayer graphene, Bernal AB-stacking graphene which is more stable than AA-stacking one
shows a metallic behavior with chiral parabolic dispersions near
the $K$ point. For trilayer graphene, there are two kinds of stacking sequences,
i.e. Bernal AB and rhombohedral ABC stacking.
Though both of them are semi-metallic, the detailed band structures differ depending on the stacking sequence.
Accordingly, their electronic properties under external
electric fields and the optical conductivity depend on
the stacking sequence.~\cite{MAoki,CLui}
Thus, even if the interlayer interaction
is weak, the stacking structures are crucial in determining
the electronic properties of the layered materials.

Recently, double-layer graphene heterostructures (DLGHs), in which hexagonal
boron-nitride (h-BN) layers are inserted between two graphene layers,
are investigated and exhibit many interesting properties such as tunable
metal-insulator transition, good room-temperature switching ratios,
strong coulomb drag and so on.~\cite {LBritnell, LPonomarenko, RGorbachev,
KHosono, AAPikalov, SMBadalyan, REVProfumo} In
these experiments, h-BN layers are used as a tunnel barrier with
varying thickness from one to dozens of layers. It is shown that
the tunneling resistance of the heterostructure is
sufficiently high for the h-BN spacer of three atomic layers, and
becomes unacceptable low for bilayer h-BN.~\cite{RGorbachev,LBritnell2, GLee}
In these DLGHs, where the number of h-BN layers is
one, two and three, the interlayer interaction between graphene and
its adjacent h-BN layer influences on the electronic properties of
the graphene. In addition, owing to the short distance between graphene
layers, the interlayer interaction between two graphene layers
is still not negligible similar to tri- and tetra-layer graphenes.
Therefore, the electronic properties of DLGHs are different from mono- and bi-layer graphenes
on a h-BN substrate. To clarify this difference in electronic
properties, in this paper we investigate the electronic
structures of DLGHs on the basis of density functional
theory. We discuss the stability and energy band structure of the
DLGHs by changing the number of h-BN layers and the
stacking sequence. Our results show that the coupling between graphene layer
and its adjacent h-BN layer becomes most stable with Ab-stacking sequence.
It is also shown that the electronic states of DLGHs can be classified into metallic or
semiconducting by the stacking sequence. The simple orbital interaction
model qualitatively well describe the stacking sequence dependence of
the energy band structures of DLGHs.

\section{\bf Calculation Method and Simulation Model}
The calculations are performed using density functional theory based
Vienna {\it Ab initio} Simulation Package (VASP).~\cite {GKresse}
Core and valence electrons are described with the projector-augmented
wave method (PAW).~\cite{JBlchl} The local density approximation (LDA) is used to treat the
exchange-correlation functional.\cite{DCeperley} The plane-wave basis
set cutoff is 500 eV and the Brillouin zone is sampled with 13$\times$13$\times$1
Monkhorst-Pack mesh. The DLGHs investigated in this paper
have the sandwiched structures; h-BN layers are inserted as the spacer between two graphene monolayers.
We label the system as G-$n$BN-G. The number of h-BN layers $n$ is changed from one to
three. As the lattice mismatch between graphene and h-BN monolayer is
around 1.8\%, the graphene and h-BN layers can be assumed to be commensurate.
To ensure negligible interaction between periodic images of the slabs in the stacking
direction, 20\AA{} vacuum
is used. In this study, all the structures are fully optimized including
atomic coordinates and supercell. The convergence thresholds for energy
and force are 10$^{-4}$eV and 0.01eV/\AA.

\section{\bf Results and Discussions}
\begin{figure*}[htbp]
\begin{center}
\includegraphics[width=10 cm]{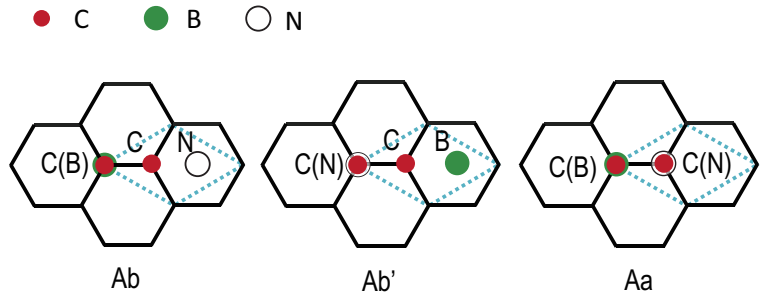}
\end{center}
\caption{Three nonequivalent configurations of graphene on a h-BN
 monolayer (Ab, Ab$'$ and Aa) seen from the top way. Carbon atoms are on
 the top layer. Boron and nitrogen atoms are on the bottom layer.
 The rhombus drawn by blue dot lines represent the primitive cells.}
\label{topview}
\end{figure*}

\begin{figure*}[htbp]
\begin{center}
\includegraphics[width=12 cm]{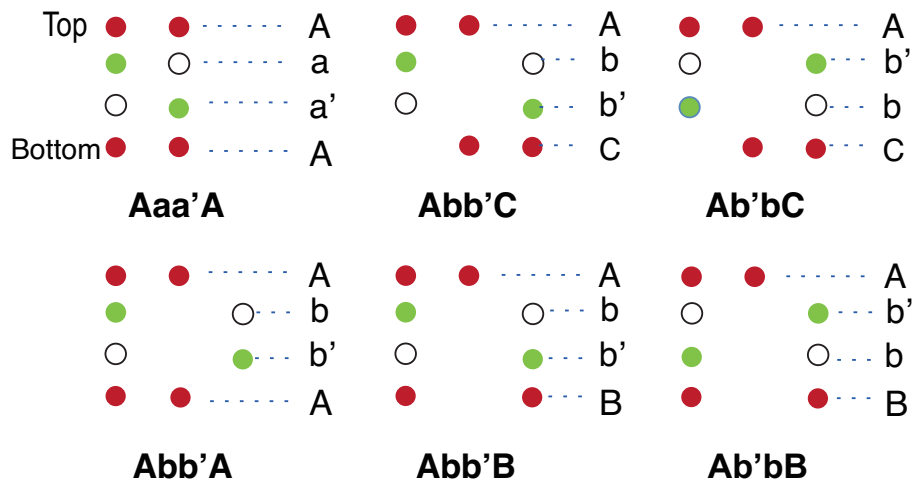}
\end{center}
\caption{Six configurations with different stacking sequences for
 G-2BN-G seen from the side way. The red, green closed circles and white open circles
mean carbon, boron, and nitrogen atoms, respectively.}
\label{structure}
\end{figure*}

\begin{figure*}[htbp]
\begin{center}
\includegraphics[width=13 cm]{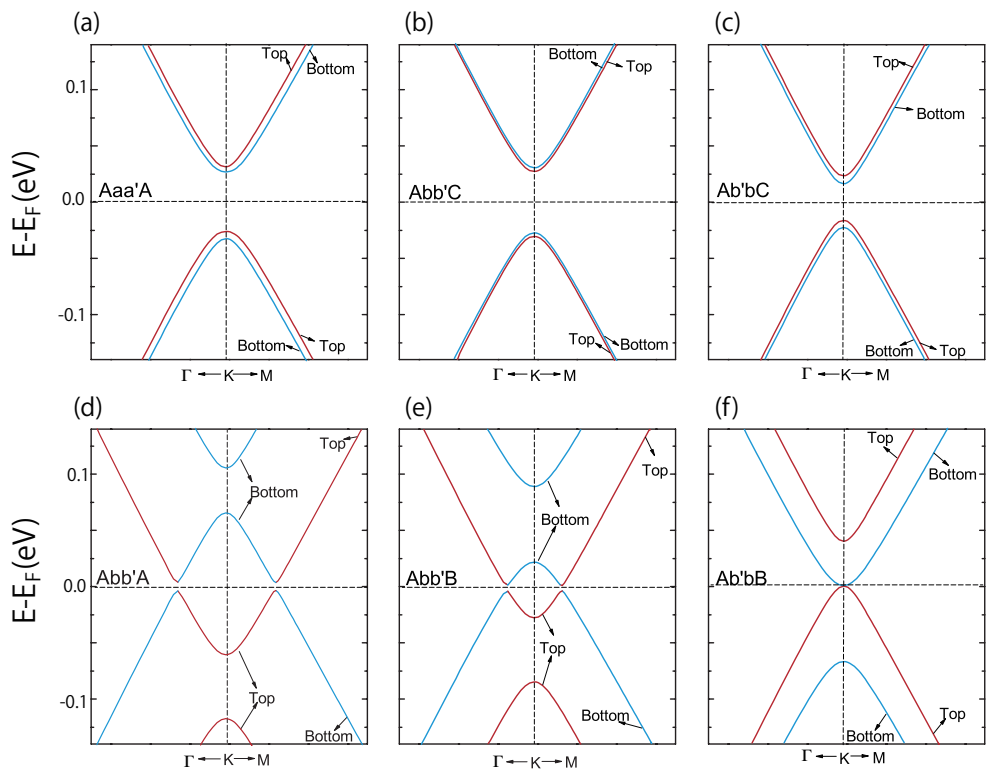}
\end{center}
\caption{Energy band structures of G-2BN-G near the Fermi level
 for the different stacking sequences.
The energy bands plotted by red (blue) lines are mainly contributed by the carbon atoms in the top (bottom)
 graphene.}
\label{band_g2bn_g}
\end{figure*}

In this work, the stacking sequences are distinguished
using the naming convention defined in the paper of Sakai {\it et
al}.~\cite {YSakai1, YSakai2} The capital letters (A, B, C) and small
letters (a, b, c) represent the relative stacking positions of graphene
and h-BN layer, respectively. For example, the Bernal stacking sequence of bilayer
graphene is labeled as AB, and the stacking form of bulk h-BN is
aa$'$. The prime symbol represents exchange of the positions of boron
and nitrogen atoms in the adjacent h-BN layers. Figure \ref{topview} shows three
nonequivalent orientations of graphene on h-BN monolayer (Ab, Ab$'$ and
Aa). Since h-BN layers always have an aa$'$ stacking,
a DLGH can have six different stacking sequences with a fixed thickness.
Figure \ref{structure} shows the six different stacking
configurations of G-2BN-G.
The corresponding energy band structures are shown in Fig.\ref{band_g2bn_g}.

Let us try to find out the most stable stacking sequence of DLGHs.
In order to clarify the relative stability among different
stacking sequences for a fixed thickness, we have evaluated the
relative differences of the total energy $\Delta$E for each stacking
configuration shown in Table \ref{table_ene_diff}. $\Delta$E is defined as the energy
difference of the total energy for each configuration measured from that
of the most stable structure for a fixed thickness of DLGH. We can see
that the most stable structure in each thickness group favor Ab-stacking between
graphene and its adjacent h-BN layer except G-2BN-G. Ab-stacking sequence is that the
carbon atoms on one of two sublattices are above boron atoms and carbon
atoms on the other sublattice are above the centers of the BN hexagonal rings.
According to this definition, b$'$C-stacking is equivalent to
bA-stacking. Therefore, the most stable configuration in G-2BN-G is
Abb$'$C, where both layers of graphene are also Ab-stacking with their
adjacent h-BN layers. The DLGHs with Aa or Aa$'$ stacking sequence,
which are least stable, have 40meV higher total energy than that of
the most stable structure.

The stability of DLGHs can also be embodied by the
interlayer distance between graphene and its adjacent h-BN layer. We
find that the average interlayer distances between the graphene and the
h-BN layer are 3.24\AA{} for Ab-stacking, 3.46\AA{} for Ab$'$-stacking and 3.52\AA{}
for Aa-stacking, respectively. This fact means that the shorter
interlayer distance gives the larger stability of DLGHs
due to the stronger interaction. Since Ab-stacking is most stable, the interaction between boron
and carbon atoms is favorable. This fact is
consistent with the results of the graphene/h-BN bilayer
superlattices.~\cite {YSakai1,YSakai2}

The energy band gaps of G-{$n$}BN-G are shown in Table \ref{table_ene_gap}.
First, we discuss the band structures of G-2BN-G and G-3BN-G. In these two cases,
the band gaps can be classified into two types:
(1) larger band gaps of dozens of meV and (2) very tiny band gaps of less than 10meV.
To clarify the mechanism of two types of the band gaps, we investigate the effect
of interlayer interaction and stacking sequence on the band structures of heterostructures.
In the DLGHs, the two graphene layers are separated by
h-BN layers, therefore graphene interacts with h-BN directly.
Previous studies showed that this interaction causes
the difference of on-site energies for
carbon atoms between two sublattices, which
induces a small band gap of dozens of meV.~\cite{EKan,NKharche,GGiovannetti}
Since G-1BN-G has much stronger interlayer interaction between graphenes than other cases,
its electronic band structure will be discussed later.

In addition to the interlayer interaction, the stacking sequences
crucially affect the electronic states near the Fermi energy.
Let us consider the configuration Abb$'$A as an example (see Fig.~\ref{structure}).
The corresponding energy band structure is shown in
Fig.~\ref{band_g2bn_g}(d). As for the top (bottom) graphene layer, the carbon
atoms on two different sublattices are located above (below) the boron (nitrogen) atoms and
the centers of hexagonal h-BN rings, respectively.
Therefore, the electrostatic
potential of the top and bottom graphene layers in Abb$'$A
heterostructure is different. In accordance with this fact, the energy
dispersion related to the top (bottom) graphene layer shifts about 0.07
eV downward (upward) as can be seen in Fig. \ref{band_g2bn_g}(d).
On the contrary, for Abb$'$C, as the carbon atoms on two different sublattices
in top (bottom) graphene layer are located above (under) the
boron atoms and centers of h-BN hexagonal rings, both the layers of graphene
have same electrostatic potential. Therefore, the energy dispersion
related to the top and bottom graphene layers overlaps at almost same
place as can be seen in Fig.\ref{band_g2bn_g}(b).

\begin{figure*}[htbp]
\begin{center}
\includegraphics[width=8 cm]{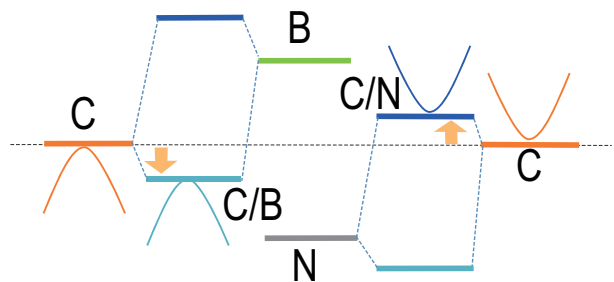}
\end{center}
\caption{Schematic energy diagram demonstrating the shift of the
 carbon energy levels after interacting with the boron/nitrogen atoms in the adjacent h-BN layer.}
\label{energy_shift}
\end{figure*}

To understand the effect of different
stacking sequences on energy band structure of DLGH,
an orbital interaction model is employed.~\cite {RQuhe} Here we just
consider the interaction between the $\pi$ orbitals of adjacent
atoms. According to atom electronegativity, i.e.
N$>$C$>$B, the energy levels of the orbitals of N,
C, B atoms (E$_{N}$, E$_{C}$, E$_{B}$) are
E$_{B}$$>$E$_{C}$$>$E$_{N}$. In Fig.\ref{energy_shift}, schematic
energy diagram is shown to illustrate the shift of energy levels induced
by the interaction between graphene and h-BN layer. For h-BN monolayer,
at $K$ point, the top of valence band is mainly contributed by
nitrogen atoms and the bottom of conduction band is contributed by boron
atoms, respectively. Therefore, when the carbon atom is located above or
below the boron atom, the occupied carbon orbital interacts with the
empty boron orbital, leading to the downward shift of the bonding state
from the original carbon level. As for the interaction between carbon
and nitrogen atoms, the situation is opposite. The empty carbon orbital
interacts with nitrogen orbital,
resulting in the upward shift of energy level of anti-bonding state.

The shift of orbital energy for carbon atom can explain the band structures
of DLGHs. In the case of tetralayer heterostructures, for
Aaa$'$A, Abb$'$C, Ab$'$bC configurations, the top and bottom graphene
layers have the same stacking sequence with their adjacent h-BN layers.
Therefore, the interaction with the adjacent h-BN layers makes the
carbon energy levels move in the same direction.
In this situation, as shown in Figs.\ref{band_g2bn_g}(a)-(c), the energy band
structures of those heterostructures are similar to that of graphene monolayer on
h-BN layer. For another three heterostructures Abb$'$A, Abb$'$B and
Ab$'$bB, the stacking of the top and bottom graphene layers with h-BN
layers are in different ways, which induces the carbon orbital energies of
two graphenes shift towards opposite directions, as shown in
Figs.\ref{band_g2bn_g}(d)-(f).
Let us take Abb$'$A as an example. The energy band
corresponding to bottom graphene moves upwards and the energy band
corresponding to top graphene moves downwards. Around the Fermi level,
the energy bands from top and bottom graphenes intersect. At the crossing point, there are
small openings of energy band gap. For Abb$'$A, it is 8meV and for Abb$'$B, it is 5meV.
We attribute the band gap opening at the crossing points to the coupling of
the top and bottom graphene layers mediated by h-BN layers. For
penta layer heterostructures Abb$'$bB and Abb$'$bC where three layers of
h-BN are inserted, the band structures are similar to that of Abb$'$A and
Abb$'$B tetralayer heterostructures. However, the
band gaps at the cross points are almost zero, which indicates the much weakened
inter-graphene interaction.

\begin{figure*}[htbp]
\begin{center}
\includegraphics[width=13 cm]{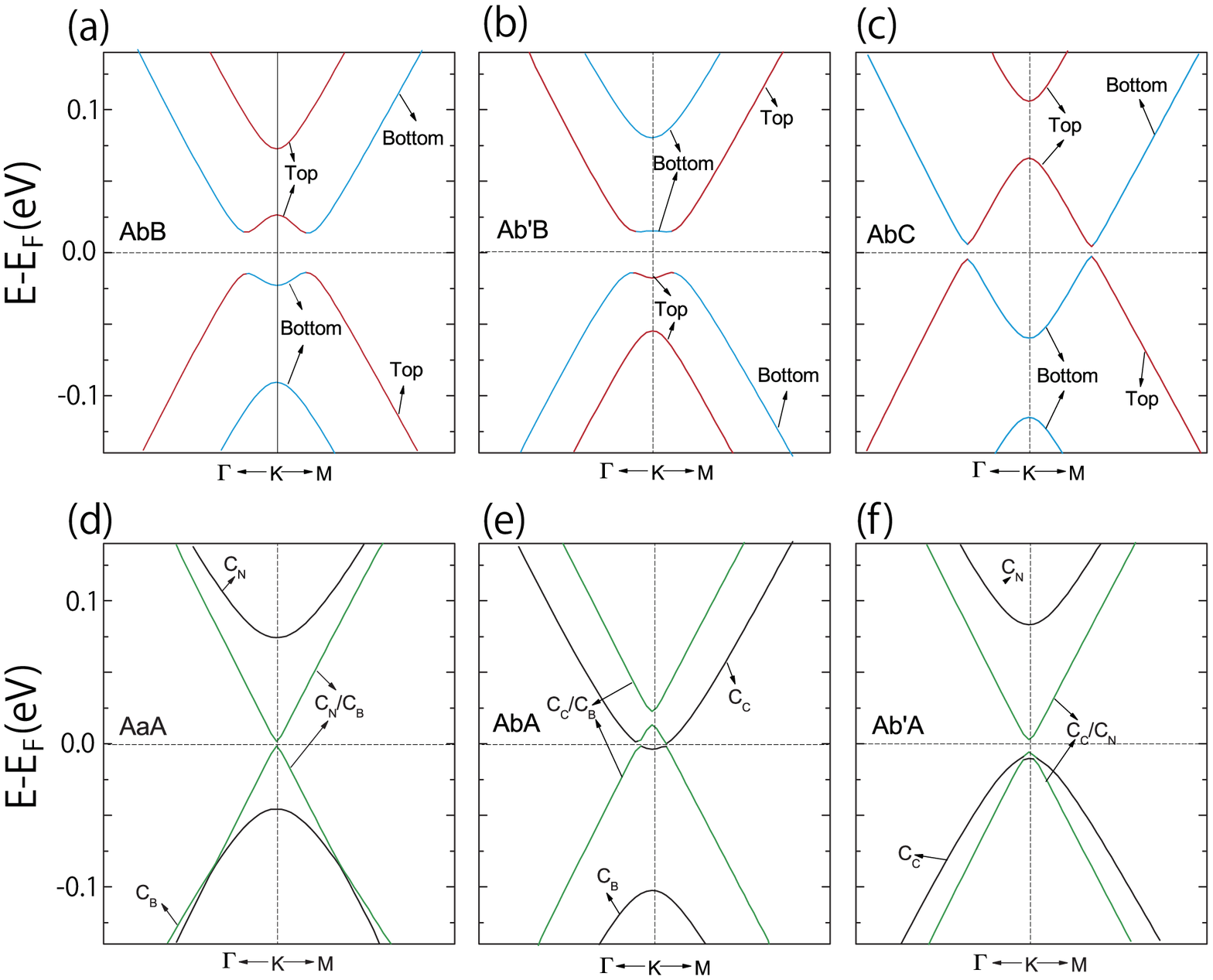}
\end{center}
\caption{Energy band structures of the six heterostructures for G-1BN-G
 near the Fermi level. In (a)-(c), the energy bands plotted by
 red (blue) lines are mainly contributed by the carbon atoms in the
 top (bottom) graphene. In (d)-(f), C$_B$, C$_N$, C$_C$ means the energy
 bands mainly contributed by the carbon atom above/below the boron,
 nitrogen or the center of hexagon of the adjacent h-BN layer. The
 green lines indicate the linear energy bands.}
\label{band_g1bn_g}
\end{figure*}

Figure \ref{band_g1bn_g} shows the energy band structures of
G-1BN-G for six different stacking sequences.
For G-1BN-G, the coupling between top and bottom graphene layers
is stronger than that of G-2BN-G and G-3BN-G. As a result, the band
structures near the Fermi level become more complex. For AbB, Ab$'$B and AbC
configurations, their energy band structures become similar to those of
Abb$'$bB, Ab$'$bb$'$B and Abb$'$bC, because of the similar stacking configuration
between graphene and its adjacent h-BN layer.
The only difference is that the interlayer distance between two graphenes is
shorter in trilayer heterostructures, resulting in the larger band gaps owing to
the stronger interlayer interaction.
For AbA, Ab$'$A and AaA configurations, though the stacking sequences between graphene
and h-BN are same as those of Abb$'$bA, Ab$'$bb$'$A and Aaa$'$aA, the band structures
are totally different. As shown in Figs.\ref{band_g1bn_g}(d)-(f), there are two
kinds of energy bands around the Fermi level. One energy band is
approximately linear at {$K$} point, and the other one is
non-linear. We find that this linear energy band is contributed by all four
carbon atoms in the unit cell, which is similar to the linear band state existing in
ABA-stacking trilayer graphene.\cite {MAoki}. For ABA-stacking trilayer graphene, the linear band
states are referred to be the bonding state of orbitals at both ends via
the second neighbor hopping parameter $\gamma_{5}$. For AbA, Ab$'$A and
AaA heterostructures, the situation is similar. Just the sandwiched
layer is changed from graphene to h-BN, so the second neighbor hopping
parameter is different. As for the nonlinear energy band,
the energy levels of the carbon atoms with same electrostatic potential
move towards same direction, because of which they compose one energy band,
labeled as C$_{B}$, C$_{N}$ and C$_{C}$.
This indicates that the coupling between top and bottom graphene plays a more important
role.

From the above results, it is clear that the band structures can
demonstrate the difference of the coupling of the two graphene
layers. The coupling of the graphene layers is strong in trilayer
heterostructures in spite of the existence of tunnel barrier of h-BN. As
the number of h-BN layers increases, the coupling is weakened.

\section{\bf Conclusions}
In this paper, first-principles calculation has been performed to
investigate the stability and the energy band structure of DLGHs,
where h-BN layers are inserted between two graphene monolayers.
Our results show that the stability and band structures
are sensitive to the stacking sequence of DLGHs.
It is found that the stability of DLGHs is mainly dependent
on the stacking sequence of graphene and its adjacent h-BN layer.
Among the three stacking sequences, Ab-stacking is most stable. In
addition, it is found that the interaction between graphene and h-BN
affects the electronic properties in two aspects. First one is modification of on-site energy
of carbon atoms between two nonequivalent sublattices in the same
graphene. The second one is shifting the band structures of the top and
bottom graphene layers, like graphene under an external electronic
field. Our results give the fundamental aspects of the electronic states
of DLGHs for the application of electronic devices.

\acknowledgment
K. W. acknowledges the financial support by Grant-in-Aid for Scientific
Research from MEXT and JSPS (Nos. 25107005, 23310083 and 20001006).

\newpage\clearpage

\begin{table}
\caption{Relative differences of total energy $\Delta$E for G-nBN-G (n=1-3).
$\Delta$E is defined as a energy difference measured from the energy of the most
stable structure for a fixed n. The structure with $\Delta$E=0 corresponds
 to the most stable one.}
\label{table_ene_diff}
\begin{tabular}{llllll}
\Hline
\multicolumn{1}{c}{G-1BN-G} & \multicolumn{1}{c}{$\Delta$E(meV)} & \multicolumn{1}{c}{ G-2BN-G} & \multicolumn{1}{c}{$\Delta$E(meV)}& \multicolumn{1}{c}{G-3BN-G} & \multicolumn{1}{c}{$\Delta$E(meV)} \\
\Hline
AbA    &  0   & Abb$'$C & 0    & Abb$'$bA    & 0   \\
AbC    & 19.1 & Abb$'$A & 18.7 & Abb$'$bC    & 18.6\\
AbB    & 22.4 & Abb$'$B & 21.9 & Abb$'$bB    & 22.0\\
Ab$'$A & 29.0 & Ab$'$bC & 38.1 & Ab$'$bb$'$A & 37.4\\
Ab$'$B & 41.3 & Ab$'$bB & 42.0 & Ab$'$bb$'$B & 41.4 \\
AaA    & 44.8 & Aaa$'$A & 44.1 & Aaa$'$aA    & 45.1\\
\Hline
\end{tabular}
\end{table}

\begin{table}[tbp]
\caption{Energy band gaps of G-nBN-G (n=1-3).}
\label{table_ene_gap}
\begin{tabular}{llllll}
\Hline
\multicolumn{1}{c}{G-1BN-G} & \multicolumn{1}{c}{GAP(meV)} & \multicolumn{1}{c}{ G-2BN-G} & \multicolumn{1}{c}{GAP(meV)}& \multicolumn{1}{c}{G-3BN-G} & \multicolumn{1}{c}{GAP(meV)} \\
\Hline
AbA    &  2.4  & Abb$'$C & 55   & Abb$'$bA    & 51 \\
AbC    &  6    & Abb$'$A & 8    & Abb$'$bC    & 0  \\
AbB    &  28   & Abb$'$B & 5    & Abb$'$bB    & 1  \\
Ab$'$A &  9    & Ab$'$bC & 37   & Ab$'$bb$'$A & 41 \\
Ab$'$B &  29   & Ab$'$bB & 1    & Ab$'$bb$'$B & 1  \\
AaA    &  3    & Aaa$'$A & 52   & Aaa$'$aA    & 61 \\
\Hline
\end{tabular}
\end{table}

\newpage\clearpage

\begin{thebibliography}{9}
\bibitem{KSNovoselov1}
K. S. Novoselov, A. K. Geim, S. V. Morozov, D. Jiang, M. I. Katsnelson,
I. V. S. Grigorieva, V. Dubonos and A. A. Firsov, Nature {\bf 438},
197 (2005).
\bibitem {YBZhang1}
Y. B. Zhang, Y. W. Tan, H. L. Stormer and P. Kim, Nature {\bf 438}, 201 (2005).
\bibitem {AKGeim1}
A. K. Geim, Science {\bf 324}, 1530 (2009).
\bibitem {KIBolotin}
K. I. Bolotin, K. J. Sikes, Z. Jiang, M. Klima, G. Fudenberg, J. Hone,
     P. Kim and H. L. Stomer, Solid State Commun. {\bf 146}, 351 (2008).
\bibitem {AABalandin}
A. A. Balandin, S. Ghosh, W. Z. Bao, I. Calizo, D. Teweldebrhan, F. Miao,
     and C. N. Lau, Nano Lett. {\bf 8}, 902 (2008).
\bibitem {TAndo}
T. Ando, J. Phys. Soc. Jpn. {\bf 74}, 777 (2005).
\bibitem {FSchweirz}
F. Schweirz, Nat. Nanotechenol. {\bf 5}, 487 (2010).
\bibitem {MAoki}
M. Aoki and H. Amawashi, Solid. State. Comm. {\bf 142}, 123 (2007).
\bibitem {MKlintenberg}
M. Klintenberg, S. Leb$\grave{e}$gue, C. Oritz, B. Sanyal, J. Fransson
     and O. Eriksson, J. Phys.: Condens. Matter {\bf 21}, 335502 (2009).
\bibitem {JYan}
J. A. Yan, W. Y. Ruan and M. Y. Chou, Phys. Rev. B {\bf 83}, 245418 (2011).
\bibitem {FZhang}
F. Zhang, B. Sahu, H. Min, and A. H. MacDonald, Phys. Rev. B {\bf 82}, 035409 (2010).
\bibitem {DHXu}
D. H. Xu, J. Yuan, Z. J. Yao, Y. Zhou and F. C. Zhang, Phys. Rev. B {\bf 86}, 201404(R) (2012).
\bibitem {CLui}
C. H. Lui, Z. Q. Li, K. F. Mak, E. Cappelluti and T. F. Heinz, Nature
     Phys. {\bf 7}, 944 (2011).
\bibitem {LBritnell}
L. Britnell, R. V. Gorbachev, R. Jalil, B. D. Belle, F. Schedin,
     A. Mishchenko, T. Georgiou, M. I. Katsnelson, L. Eaves,
     S. V. Morozov, N. M. R. peres, J. Leist, A. K. Geim,
     K. S. Novoselov and L. A. Ponomarenko, Science, {\bf 335}, 947
     (2012).
\bibitem {LPonomarenko}
L. A. Ponomarenko, A. K. Geim, A. A. Zhukov, R. Jalil, S. V. Morozov,
     K. S. Novoselov, I. V. Grigorieva, E. H. Hill, V. V. Cheianov,
V. I. Fal$'$ko, K. Watanabe, T. Taniguchi and R. V. Gorbachev, Nature
     Phys. {\bf 7}, 958 (2011).
\bibitem {RGorbachev}
R. V. Gorbachev, A. K. Geim, M. I. Katsnelson, K. S. Novoselov,
     T. Tudorovshiy, I. V. Grigorieva, A. H. MacDonald,
     S. V. Morozov,  K. Watanabe, T. Taniguchi and
     L. A. Ponomarenko, Nature Phys. {\bf 8}, 896 (2012).
\bibitem {KHosono}
K. Hosono and K. Wakabayashi, Appl. Phys. Lett. {\bf 103}, 033102 (2013).
\bibitem {AAPikalov}
A. A. Pikalov and D. V. Fil, Nanoscal Res. Lett. {\bf 7}, 145 (2012).
\bibitem {SMBadalyan}
S. M. Badalyan and F. M. Peeters, Phys. Rev. B {\bf 85}, 195444 (2012).
\bibitem {REVProfumo}
R. E. V. Profumo, R. Asgari, M. Polini, and A. H. MacDonald, Phys. Rev. B 85, 085443 (2012).
\bibitem {LBritnell2}
 L. Britnell, R. V. Gorbachev, R. Jalil, B. D. Belle, F. Schedin,
     M. I. Katsnelson, L.Eaves, S. V. Morozov, A. S. Mayorov,
     N. M. R. Peres, A. H. C. Neto, J. Leist, A. K. Geim,
     L. A. Ponomarenko and K. S. Novoselov, Nano Lett. {\bf 12},
     1707(2012).
\bibitem {GLee}
G. H. Lee, Y. J. Yu, C. Lee, C. Dean, K. L. Shepard, P. Kim and J. Hone,
     Appl. Phys. Lett. {\bf 99}, 243114 (2011).
\bibitem {GKresse}
G. Kresse and J.Hafner, Phys. Rev. B {\bf 54}, 11169 (1996).
\bibitem {JBlchl}
J. P. Perdew, Phys. Rev. B {\bf 50}, 17953 (1994).
\bibitem {DCeperley}
D. M. Ceperley and B. J. Alder, Phys. Rev. Lett. {\bf 45}, 566 (1980).
\bibitem {YSakai1}
Y. Sakai, and S. Saito, J. Phys. Soc. Jpn. {\bf 81}, 103701 (2012).
\bibitem {YSakai2}
Y. Sakai, T. Koretsune, and S. Saito, Phys. Rev. B {\bf 83}, 205434 (2011).
\bibitem {EKan}
E. Kan, H. Ren, F. Wu, Z. Li, R. F. Lu, C. Y. Xiao, K. M. Deng and
     J. L. Yang, J. Phys. Chem. C {\bf 116}, 3142 (2012).
\bibitem {NKharche}
N. Kharche and S. K. Nayak, Nano Lett. {\bf 11}, 5274 (2011).
\bibitem {GGiovannetti}
G. Giovannetti, P. A. Khomyakov, G. Brocks, P. J. Kelly and J. Brink,
     Phys. Rev. B {\bf 76}, 073103 (2007).
\bibitem {RQuhe}
R. Quhe, J. X. Zheng, G. F. Luo, Q. H. Liu, R. Qin, J. Zhou, D. P. Yu, S. Nagase, W. N. Mei, Z. X. Gao and J. Lu,
NPG Asia Materials {\bf 4}, e6 (2012).
\end{thebibliography}
\end{document}